# Edge-Epitaxial Growth of InSe Nanowires toward High-Performance Photodetectors

Song Hao, Shengnan Yan, Yang Wang, Tao Xu, Hui Zhang, Xin Cong, Lingfei Li, Xiaowei Liu, Tianjun Cao, Anyuan Gao, Lili Zhang, Lanxin Jia, Mingsheng Long, Weida Hu, Xiaomu Wang, Pingheng Tan, Litao Sun, Xinyi Cui, Shi-Jun Liang* and Feng Miao*

Dr. S. Hao, S. N. Yan, X. W. Liu, T. J. Cao, A. Y. Gao, L. L. Zhang, L. X. Jia, Dr. S. J. Liang and Prof. F. Miao
[1]National Laboratory of Solid-State Microstructures, School of Physics, Collaborative Innovation Center of Advanced Microstructures, Nanjing University, Nanjing 210093, China
Corresponding authors: miao@nju.edu.cn; sjliang@nju.edu.cn
Y. Wang, Dr. M. S. Long, Prof. W. D. Hu
[2]State Key Laboratory of Infrared Physics, Shanghai Institute of Technical Physics, Chse Academy of Sciences, Shanghai 200083, China
Dr. T. Xu, H. Zhang, Prof. L. T. Sun
[3]SEU-FEI Nano-Pico Center, Key Laboratory of MEMS of Ministry of Education, Southeast University, Nanjing 210096, China
X. Cong, Prof. P. H. Tan
[4]State Key Laboratory of Superlattices and Microstructures, Institute of Semiconductors, College of Materials Science and Opto-Electronic Technology, Chinese Academy of Sciences, Beijing 100083, China
L. F. Li, Prof. X. M. Wang
[5]School of Electronic Science and Technology, Nanjing University, Nanjing 210093, China
Prof. X. Cui
[6]State Key Laboratory of Pollution Control and Resource Reuse, School of the Environment, Nanjing University, Nanjing 210046, China

**Abstract:** Semiconducting nanowires offer many opportunities for electronic and optoelectronic device applications due to their special geometries and unique physical properties. However, it has been challenging to synthesize semiconducting nanowires directly on $SiO_2$/Si substrate due to lattice mismatch. Here, we developed a catalysis-free approach to achieve direct synthesis of long and straight InSe nanowires on $SiO_2$/Si substrate through edge-homoepitaxial growth. We further achieved parallel InSe nanowires on $SiO_2$/Si substrate through controlling growth conditions. We attributed the underlying growth mechanism to selenium self-driven vapor-liquid-solid process, which is distinct from conventional metal-catalytic vapor-liquid-solid method widely used for growing Si and III-V nanowires. Furthermore, we demonstrated that the as-grown InSe nanowire-based visible light photodetector simultaneously possesses an extraordinary photoresponsivity of 271 A/W, ultrahigh detectivity of $1.57 \times 10^{14}$ Jones and a fast response speed of microsecond scale. The excellent performance of the photodetector indicates that as-grown InSe nanowires are promising in future optoelectronic applications. More importantly, the proposed edge-homoepitaxial approach may open up a novel avenue for direct synthesis of semiconducting nanowire arrays on $SiO_2$/Si substrate.

Introduction

**INTRODUCTION**

With high mobility, direct band gap and good stability, two-dimensional (2D) Indium Selenide (InSe) has emerged as one of the most promising candidates for the next generation electronic [1-3] and optoelectronic devices [4-6]. Reducing the dimension of 2D InSe to one-dimensional (1D) InSe nanowire can enhance the density of states and further lead to stronger photoresponse, making it more promising in optoelectronic devices applications. [7, 8] Recent works have predicted that 1D InSe nanoribbon could exhibit tunable half-metallicity and intrinsic ferromagnetic properties,[9, 10] which may further diversify its device applications. Nevertheless, there has been no reports on the growth of InSe nanowire so far, except chemical vapor deposition (CVD) growth of InSe microflakes and films. [11-13] It is also highly required to synthesize InSe nanowires directly on $SiO_2$/Si substrate, to keep compatible with the planar semiconductor technology currently used for industry.

Vapor-solid-solid (VSS) and vapor-liquid-solid (VLS) are two conventional approaches used for synthesizing nanowires. However, the nanowires grown on $SiO_2$/Si substrate through VSS approach usually suffer from issues of inhomogeneity and random orientation due to spontaneous nucleation and growth of nanowires. These issues lead to a challenge in integrating nanowires into electronic circuits with standard planar processing technology. [7, 14, 15] The issues faced by the VSS approach can be overcome by metal catalysts-assisted VLS approach. [16-20], which is the most widely used avenue of synthesizing semiconducting nanowires. However, the use of metal catalysts *e.g.* gold and silver, would give rise to many recombination/generation centers, [21, 22] *via* creating deep-level defects in nanowires. These defects would affect electrical and optical properties of as-grown nanowires and induce detrimental effects on the performance of electronic and optoelectronic devices based on the nanowires. Furthermore, growth of high-quality semiconducting nanowire planar arrays through VLS approach requires the use of single-crystalline substrate, such as sapphire, GaN and SiC, which is not only expensive but also usually limited by lattice matching issue. [23-25] Therefore, these approaches are not suitable for synthesis of single-crystal and well-aligned

InSe nanowires. It is highly desirable to develop a method to grow planar InSe nanowires, which is a catalyst-free and compatible with CMOS manufacturing processing.

In this article, we for the first time report the synthesis of parallel single-crystal InSe nanowires directly on $SiO_2$/Si substrates through a catalysis-free edge homoepitaxial approach. Through the detailed characterization of tunneling electron microscopy, X-ray photoelectron spectroscopy, Raman spectrum and atomic force microscopy, we propose that selenium self-driven vapor-liquid-solid process governs the underlying growth of InSe nanowires. The measurement of Energy-dispersive spectroscopy indicates that InSe molecules adsorbed on edges of InSe microflakes were decomposed into the selenium droplets and served as non-metal catalyst, and eventually driven the synthesis of parallel nanowires by edge-epitaxial vapor-liquid-solid process. The as-synthesized InSe nanowire-based field effect transistors exhibit typical n-type transport characteristics. Furthermore, we show that the InSe nanowire-based photodetectors have excellent performance with detectivity of $1.57 \times 10^{14}$ Jones and response speed of microsecond time scale. Our work may pave the way towards the large-scale synthesis of InSe nanowire arrays on $SiO_2$/Si substrates and the realization of high-performance optoelectronic devices.

**RESULTS AND DISCUSSION**

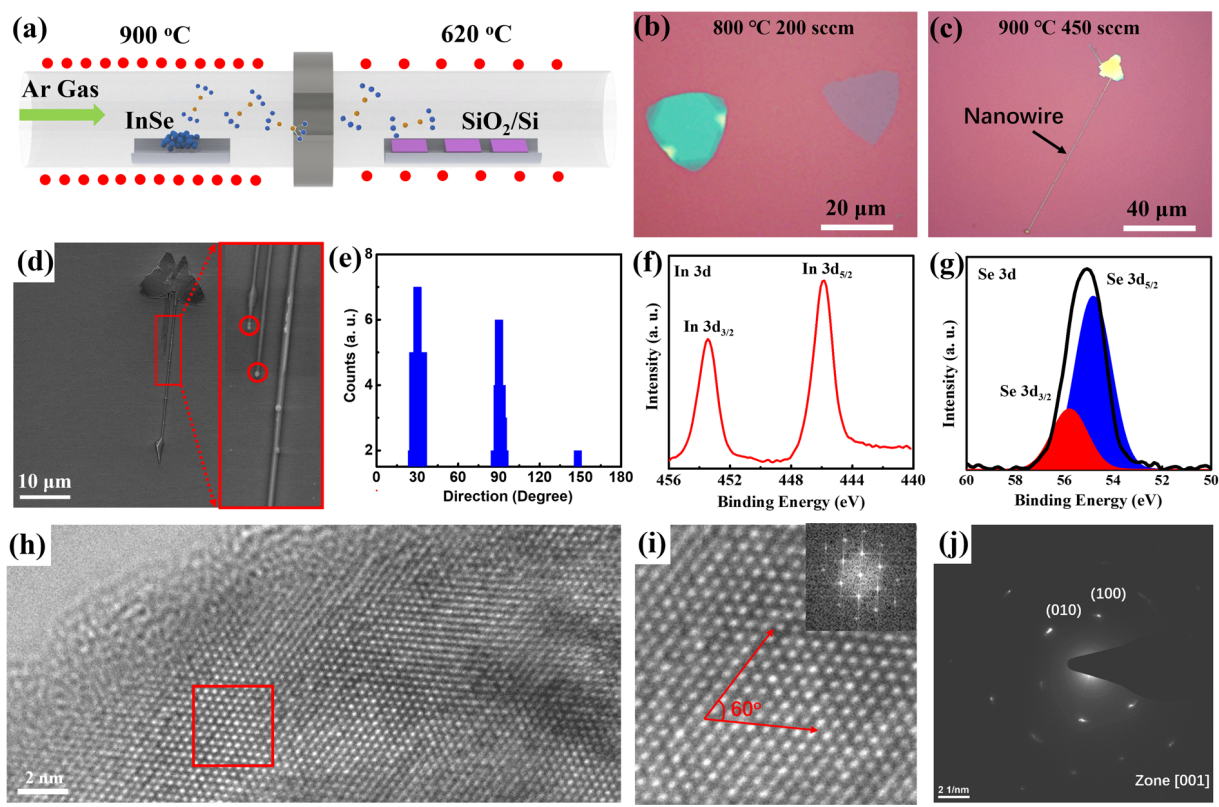

**Figure 1. Direct synthesis of InSe nanowires on SiO₂/Si substrate in a low-pressure CVD furnace.** (a) A schematic diagram of InSe microflakes and nanowires synthesized on SiO₂/Si substrates with InSe powders as precursors. (b, c) The optical microscopy images of InSe microflakes and nanowires, which were synthesized at distinct temperatures and flow rates of carrier gas, respectively. (d) The SEM image of parallel InSe nanowires and high-magnification image (Inset) marked in red-square area, shows droplets at the vertex of nanowires. (e) The histogram of nanowire orientations with respect to one of edges of triangle-shape InSe microflake, indicating orientation-dependent growth behavior. (f, g) The XPS narrow scans of In 3d and Se 3d core level spectra for as-synthesized sample, respectively. High-resolution TEM image of InSe edge (h) and high-magnification TEM image (i) and corresponding Fast Fourier Transform pattern (Inset) and Selective Area Electron Diffraction pattern (j).

We synthesized nanowires by employing InSe powders as reactant precursor, as schematically shown in Figure 1a. When the growth temperature reaches up to 800 °C, we obtained InSe microflakes with triangle shape (Figure 1b), which has been characterized by the photoluminescence, Raman and EDS spectra (Figure S1). With the growth temperature rising to 900 °C, nanowires (Figure 1c) were synthesized. Note that it has been challenging to grow parallel nanowires on amorphous substrates, owing to random distribution of nucleus. We used Scanning Electron Microscope to characterize the morphology of as-grown nanowires, with results shown in Figure 1d. The magnified image as marked by the red box shows that the nanowires are well aligned and that the as-grown nanowires are merged with edges of the microflake. These nanowires grew out of the edges with preferential directions (See Figure S2). This is justified by the histogram shown in Figure 1e, where the angle between nanowires and the flake edge are mostly centered at 30º and 90º. Besides, we found that the nanowires are terminated with tips marked by the red circles, similar to metal catalysis droplets used in the VLS growth of semiconducting silicon, germanium and III-IV nanowires. [14-19] It can be noticed that monodispersed flakes are regular triangle shape, while flakes merged with nanowires look irregular. The irregular edges appear due to the fact that an elevated growth temperature gives rise to higher chemical activity of species, multiple nuclei and faster growth velocity. [26-28] It is worth to point out that the InSe nanowires are grown on $SiO_2$/Si substrate without any metal catalysis in this work. Furthermore, we performed XPS measurements of as-synthesized samples to identify the chemical elements (Figures 1f and 1g). After calibrating with C 1s peak at 284.5 eV, the doublet peaks are actually located at ~ 453.4 eV and ~ 445.9 eV below Fermi level are assigned to In $3d_{3/2}$ and $3d_{5/2}$ core levels, in accordance with InSe flakes. [29-31] The deconvoluted doublet peaks located at ~55.8 eV and ~54.8 eV correspond to Se $3d_{3/2}$ and $3d_{5/2}$ core levels, respectively. The oxidization of as-grown samples seems unlikely by the absence of the peak for Se-O band centered at ~59.0 eV. The difference values of In $3d_{3/2}$ and $3d_{5/2}$ peaks between our as-synthesized sample and oxidized InSe are near 2.0 eV, demonstrating that as-synthesized sample is not oxidized InSe. Moreover, the O 1s spectra results (see Figure S3) undoubtedly rule out the possibility of oxidation of as-synthesized sample since the binding energy of O 1s is distinctly different from that of air-oxidized $InSe_{1-x}O_x$. [31] With the respective integrated peak areas of the XPS spectrum, we estimated the atomic stoichiometric ratio between selenium and indium to be ~0.9, indicating the successful synthesis of InSe nanowires. This conclusion is also corroborated by the measurement of Energy-dispersive X-ray spectroscopy (EDX), as will be discussed later. To investigate the crystal structures of as-grown InSe nanowires, we further carried out measurements of transmission electron microscopy (TEM) and selective area electron diffraction (SAED). The high-resolution TEM lattice fringe (Figure 1h) and magnified TEM image (Figure 1i) exhibit periodic arrangement of lattice crossing each other at an angle of 60º. This is an indication of typical hexagonal crystal phase, which is consistent with the SEAD pattern (Figure 1j).

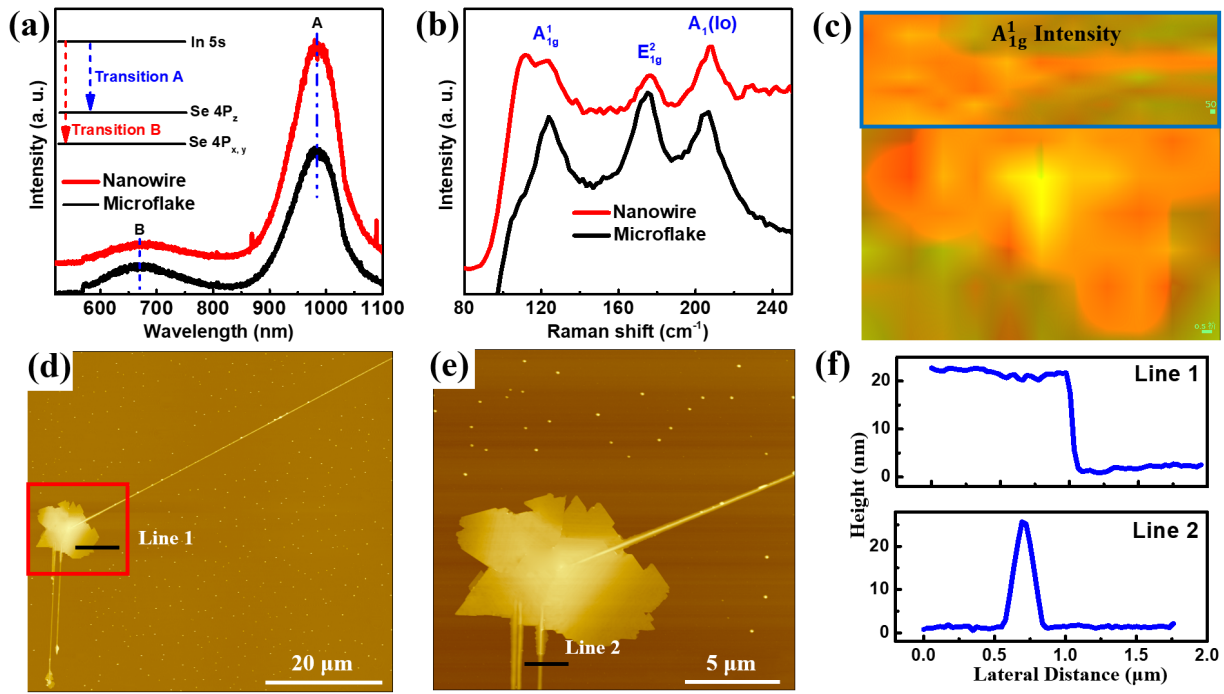

**Figure 2. Structural information of as-synthesized InSe nanowire.** (a, b) Photoluminescence and Raman spectra obtained from microflake (black circle) and nanowire (red circle) show similar characteristic peaks and the same lattice structure. (c) Raman intensity mapping of $A_{1g}^1$ characteristic mode for InSe microflake and nanowire shows identical intensity and indicate homogeneous structure. (d, e) The large scale and zoomed-in AFM topography images. (f) The corresponding height profiles extracted from black solid lines in the AFM topography images.

The structural properties of as-synthesized InSe nanowire were further characterized by Raman/Photoluminescence (PL) spectroscopy. The PL spectra (Figure 2a) of InSe nanowire (red) shows the same peak as that of microflake (black) at ~680 nm, which corresponds to the direct transition from In 5s to Se $4P_{x,y}$ state (as indicated in transition B).[8, 13, 32] The indicated transition A located at ~984 nm (1.26 eV) is attributed to the optical excitation from In 5s to Se $4P_z$ state (topmost valence band). For a bulk InSe, the PL peak can be distinguished from Si substrate (Figure S4). The optical band gap of the InSe nanowire is smaller than that of few layer InSe microflake (Figure S1). It has been reported that Raman peaks of InSe would gradually vanish along with oxidation effect.[31] However, Raman peaks of nanowires can be clearly observed in Figure 2b, indicating that our as-grown samples have not been oxidized. The Raman peaks different from the literature may be attributed to the strain-induced shift, which has been widely reported in the prior works.[33-36] Noted that the Raman spectrum of InSe nanowire exhibits an extra peak at 111.6 cm$^{-1}$, which may be resulting from the strain-induced effect. It does not arise from the fluorescence effect since the nearest photoluminescence peak located at 680 nm is far away from the Raman peak excited by laser light with 532 nm. The highly uniform contrast of $A_{1g}^1$ intensity mapping (Figure 2c) indicates homogeneous structure of the as-grown InSe nanowire. This also further justifies that nanowires homoepitaxially grow out of the edge of InSe microflake. To further reveal the structural configuration between nanowire and the edge of microflake, we performed atomic force microscopy (AFM) characterization of the as-synthesized samples. The large-scale AFM topography image (Figure 2d) reveals morphological structure similar to the results of SEM. The zoomed-in AFM topography image (Figure 2e) marked by the red-square in Figure 2d clearly shows that nanowires extend to interior of microflake rather than join at the edges of microflake, indicating that nanowires simultaneously grow with the microflake within the

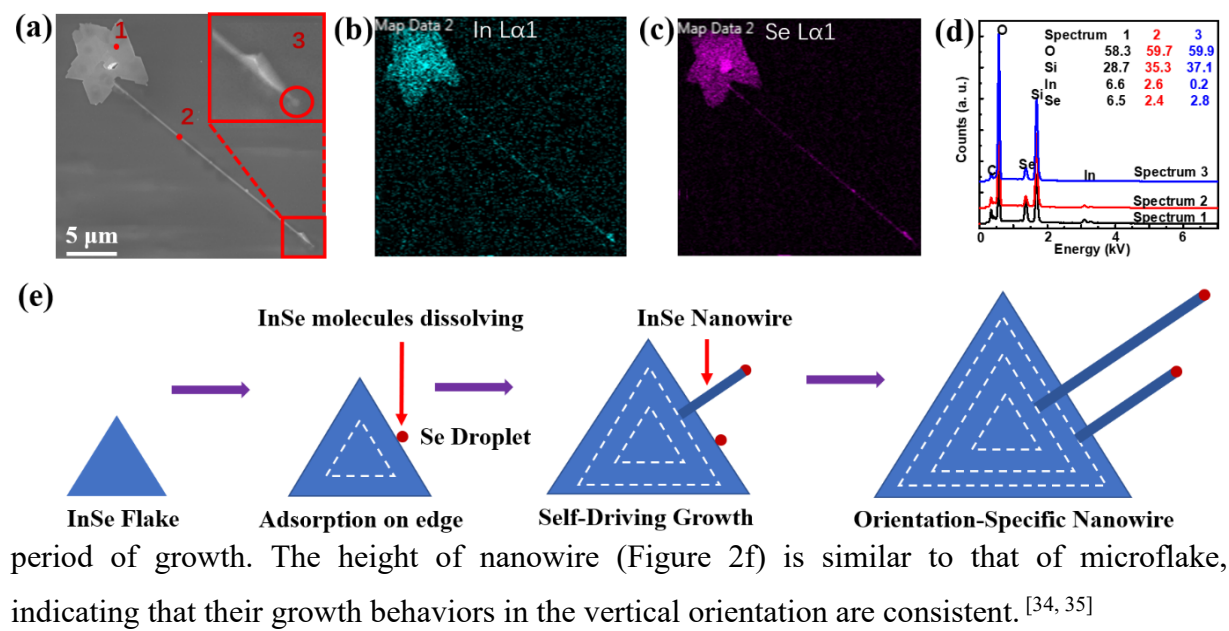

period of growth. The height of nanowire (Figure 2f) is similar to that of microflake, indicating that their growth behaviors in the vertical orientation are consistent.[34, 35]

**Figure 3. The mechanism of edge-homoepitaxial VLS growth of InSe nanowires.** (a-d) The SEM image of nanowire configuration, corresponding EDS mappings of Indium and Selenium, and elemental analysis. (e) Schematic representation of the edge-homoepitaxial VLS growth mechanism of planar InSe nanowires, which is distinct from conventional metal-catalysis VLS growth of semiconducting nanowire.

To understand the growth mechanism of the parallel nanowires, we systematically performed Energy-dispersive spectroscopy (EDS) measurements on the as-grown samples with assistance of SEM (Figures 3a-c). The identical color contrast in the EDS mappings of In (Figure 3b) and Se (Figure 3c) justifies the uniform elemental distribution in the as-grown samples. In addition, the atomic ratio of In to Se revealed via EDS spectrum (Figure 3d) is ~ 0.9, which is fairly close to its nominal stoichiometry. As shown in Figure 3a, a droplet-like point marked by the red circle is located at the vertex of nanowire. The chemical element analysis results of EDS spectrum (Figure 3d) explicitly point out that Se element is dominant over In element in this droplet-like point (*i.e.* 2.8/0.2), while other parts of as-grown samples are composed by InSe compound. All these experimental evidences shown above suggest a mechanism distinct from conventional metal catalyst-assisted VLS growth of nanowire. We propose the following mechanism responsible for the growth of parallel InSe nanowires, which is schematically shown in Figure 3i. When the growth temperature rises close to 800 °C, InSe microflakes with triangle shape are initially synthesized on the $SiO_2$/Si substrate after nucleation, similar to CVD-growth of $MoS_2$ microflakes. [37] Then, at elevated temperature of ~900 °C, InSe vapor molecules are partially decomposed into selenium droplets due to high chemical activity [7, 37] and the resulting selenium droplets are adsorbed on the edges of InSe microflakes. The InSe species dissolve into supersaturated eutectic Se droplets and precipitate at the edge of microflakes. Perfect lattice matching and low binding energy lead to the growth of InSe nanowires out of the edge of microflakes. [20] Finally, more selenium droplets crawl on the surface due to the formation of molten InSe products, yielding the planar parallel InSe nanowires. It is noteworthy that a high rate of carrier gas flow affects the growth of InSe nanowires, which guarantees a high concentration pressure during the growth process. [38]

The edge-homoepitaxial growth behavior is different from conventional growth governed by catalyst-assisted VLS process. The crystallographic orientation of VLS-governing synthesized nanowires is thermodynamically determined by the interface between the liquid droplet of metal catalyst and solid nuclei of nanowires. [16, 19] We note that such edge-induced epitaxial growth behaviors are ordinary in CVD-grown 2D heterostructures. However, edge-homoepitaxial parallel InSe nanowires have not yet been reported so far and more theoretical works are required to offer a deeper understanding of its underlying growth mechanism. [39] Compared with conventional VLS-governing synthesis of InSe nanowires (Figure S5), edge-homoepitaxial growth process enables the synthesis of high-quality and well-aligned nanowires directly on $SiO_2$/Si substrate. Therefore, this approach could be compatible with the state-of-art planar processing and pave the way for the realization of future integrated electronic and optoelectronic device applications based on nanowires. Given that the edges of 2D layered materials provide sites of high chemical reactivity, low-melting elements constituting 2D layered materials are easily adsorbed on edges to serve as catalysts. Thus, the edge-epitaxial growth mechanism may be applicable to growth of other semiconducting nanowires from III-VI layered semiconductors and semiconducting transition metal dichalcogenides. It should be pointed out that since epitaxial growth of InSe nanowire takes place at the spots along the edge of pre-grown InSe microflakes, the edge-homoepitaxy induced synthesis of parallel InSe nanowires over the whole $SiO_2$/Si substrate could be realized if pre-introducing microflake arrays or nucleation spots with specific lattice orientation. [38] For example, controlling growth orientation on mica substrate allows for synthesizing atomically-thin $In_2Se_3$ flake arrays. [40] Alternatively, via pre-defined the position of nucleation, well-defined PbS nanoplate arrays have been achieved. [41]

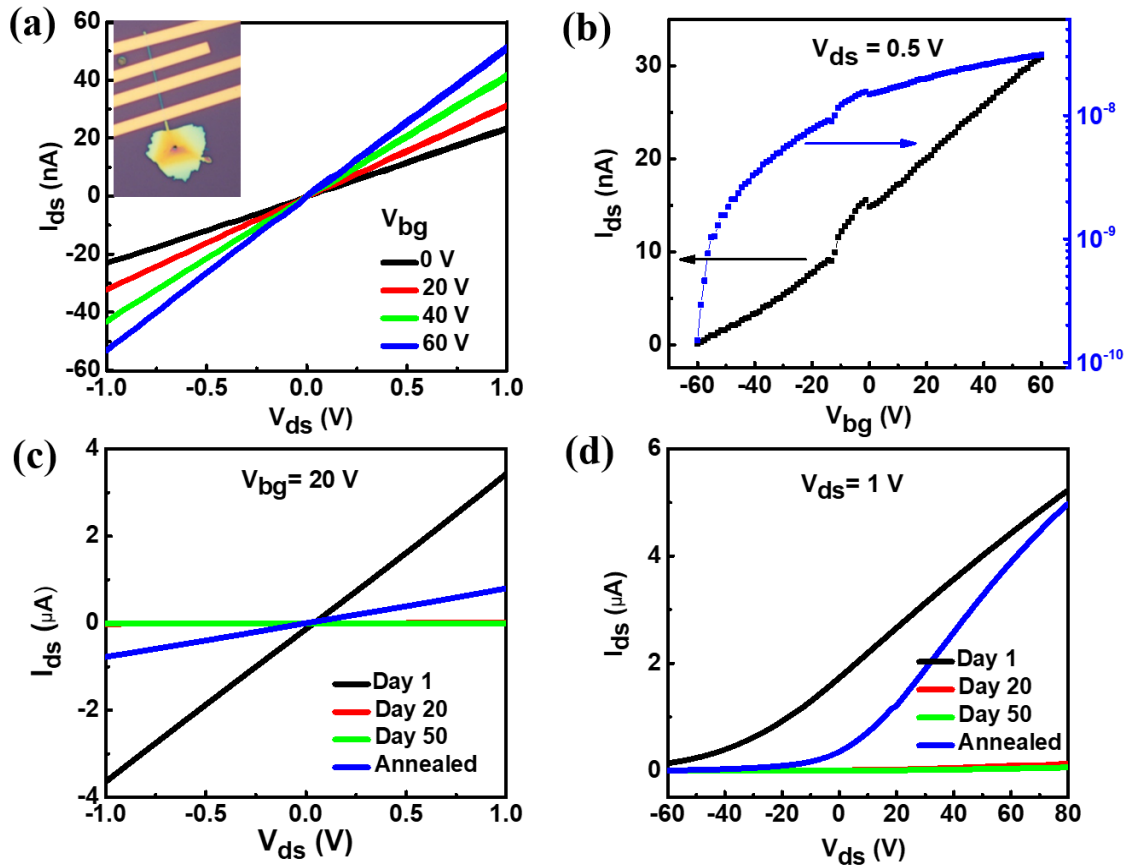

**Figure 4. Electronic properties of as-synthesized InSe nanowire.** (a) The $I_{ds}$-$V_{ds}$ characteristic curves of InSe nanowire-based transistor (Inset) with $V_{bg}$ ranging from 0 V to 60 V. The linear dependence indicates the formation of Ohmic contact. (b) The $I_{ds}$-$V_{bg}$ curves of InSe nanowire-based FET at $V_{ds}$= 0.5 V. (c, d) The time-dependent output/transfer curves of InSe based FET exposed to air and output/transfer curves after annealing in argon atmosphere, respectively.

We fabricated InSe nanowire-based field-effect transistor (FET) devices to study its electronic properties. The linear $I_{ds}$-$V_{ds}$ characteristics of InSe nanowire FET (Figure 4a) suggest the formation of Ohmic contact. Schottky contact can be realized in thin InSe nanowire devices (Figure S6), which may arise from the increased band gap in the thinner nanowires. [41] Furthermore, the transfer curves of InSe nanowire FET indicates that as-synthesized InSe nanowires are typical n-type semiconductors, similar to InSe flakes. [1, 3, 11-13] Based on the transfer curve, the extracted field-effect mobility of InSe nanowire (Figure 4b) is 1.2 cm$^2$/V·s, which is similar to that of as-grown InSe microflakes (Figure S7) but higher than that of previously reported CVD-grown InSe flakes. [11, 12] Note that all the device fabrication processes and measurements have been done without any environmental control or optimization. Considerable enhancement of carrier mobility is expected via dielectric engineering or processing materials in a glove box filled with nitrogen gas. [1, 43]

The stability of fabricated devices is essential to practical applications of materials. We evaluated the stability properties of as-synthesized InSe devices with exposure to ambient conditions. After 50 days, we did not observe obvious degradation of as-synthesized InSe samples through optical microscopy images (Figure S8), indicating the superior ambient stability over other environment-sensitive materials such black phosphorus. [44] Although performance degradation was observed (Figures 4c and d and Figure S9) possibly due to surface oxidation of InSe nanowire or physical adsorbates (such as $O_2$ and $H_2O$ molecules) on InSe surface,[45] we observed that annealing in argon atmosphere can partially recover the device performance, likely by removing the adsorbates.

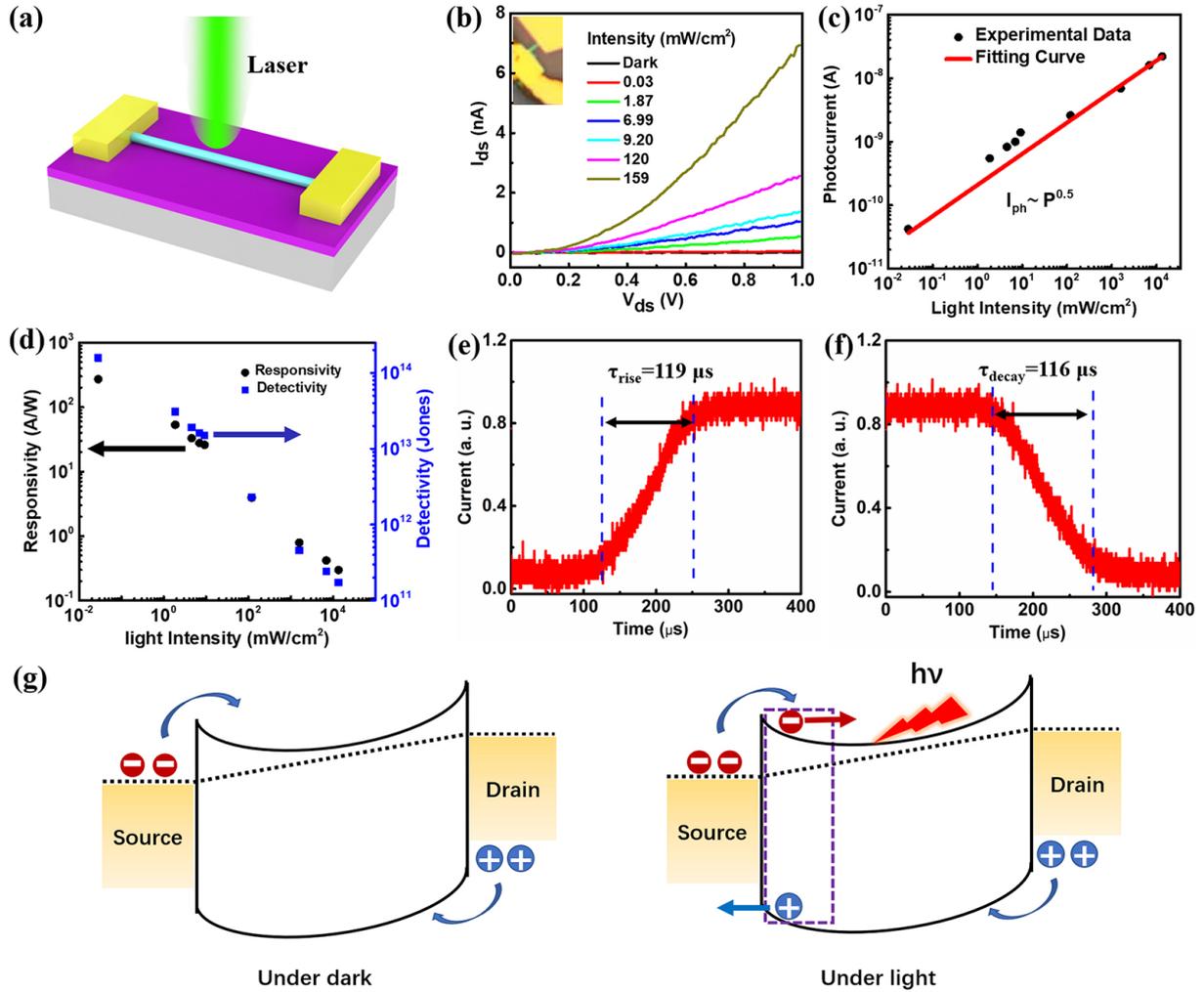

**Figure 5. Photoresponse performance of as-synthesized InSe nanowire-based photodetector.** (a) A schematic diagram of individual InSe nanowire-based visible light photodetector. (b) Output $I_{ds}$-$V_{ds}$ characteristic curves measured with and without exposure to a focused 520 nm laser beam at $V_{bg}$=0 V. The Inset is optical microscopy image of the InSe nanowire with channel width of 300 nm photodetector. (c) Measured photocurrent and fitting curve for different incident light intensities at $V_{ds}$= 1V and $V_{bg}$=0 V. (d) Photoresponsivity and detectivity for different light power intensities. (e, f) Rise and decay time of device by using a chopped laser with a frequency of 1 kHz. The rise and fall time are defined as the photocurrent increased from 10 to 90% and decreased from 90 to 10%, respectively. (g) Operating principle of the InSe-based photodetector.

Given that 1D nanowires are expected to show more competitive optoelectronic properties due to enhanced density of states, [7, 8] we fabricated individual InSe nanowire-based photodetector devices, as schematically shown in Figure 5a. The typical $I_{ds}$-$V_{ds}$ curves (Figure 5b) indicate the formation of Schottky barriers and increased $I_{ds}$ under light illumination (520 nm). The net photocurrent $I_{ph}$, defined by $|I_{light}| - |I_{dark}|$, nonlinearly increases with the light intensity P (Figure 5c). Varying trend of $I_{ph}$ versus light intensity can be fitted by a power law relation of $I_{ph} \sim P^k$ with $k=0.5$ over the range from 28.3 μW/cm$^2$ to 13450 mW/cm$^2$, indicating occurrence of complex processes in the device, *i.e.* generation and recombination of electron-hole pairs and trapping. [46]

To further evaluate the performance of InSe nanowire-based photodetectors, we extracted the photoresponsivity (*R*) and specific detectivity (*D\**) through the light intensity dependence of photocurrent. Photoresponsivity is defined as $R=I_{ph}/P \cdot A$, where *A* is the cross-sectional area of nanowires. Specific detectivity can be defined as $D^*=R \cdot A^{1/2}/(2e \cdot I_{dark})^{1/2}$, which is used to evaluate the minimum detectable signal. [46] Increasing the light intensity leads to a reduction in both *R* and *D\** (Figure 5d), which may be due to the generation and recombination of photogenerated electron-hole pairs. [13, 46] The *R* and *D\** can reach up to 271 A/W and 1.57×10$^{14}$ Jones respectively under low incident light intensity of 28.3 μW/cm$^2$, superior over these of as-synthesized InSe microflakes (Figure S10) and comparable to commercially available photodetectors. [47] Fast photoresponse is also of crucial significance to practical applications of the photodetectors based on InSe nanowire. The time-resolved photocurrent measurements reveal (Figures 5e, 5f and S11) a rise time of 119 μs and a decay time of 116 μs, respectively. As illustrated in Table 1, the specific detectivity (1.57 × 10$^{14}$ Jones) of InSe nanowire device outperforms other photodetectors based on 1D and 2D Selenides and Indium-related compounds, meanwhile its response time is comparable to mechanical exfoliated InSe microflakes with graphene as electrodes. [13, 48-50]

To shed light on the photoresponse mechanism of the InSe nanowire photodetectors, we depicted the corresponding energy band diagrams under the conditions of dark and light illumination, as shown in Figure 5g. At the dark condition, the presence of the Schottky barriers gives rise to a small current under a source and drain voltage. Upon light illumination, photo-generated electron-hole pairs in the Schottky junction region are rapidly separated by the strong built-in electric field and increase the electric conductivity, which is responsible for fast response speed of the photodetectors. [51, 52] This is justified by the experimental results of photocurrent mapping (Figure S12), indicating that the strongest photoresponse occurs near the regions of InSe/Metal electrode contact. We note that similar reports have been observed in ZnO-based photodetectors, in which Schottky junctions are deliberately formed to obtain an ultrafast response time. [51,53]

**Table 1.** Comparison of phototransistor performance based on the CVD-grown InSe nanowire with other high-performance 1D or 2D materials.

| Materials | Fabrication Method | Detectivity (Jones) | Rise time (ms) | References |
|---|---|---|---|---|
| $In_2Se_3$ Nanowire | CVD | — | 300 | [48] |
| InSe Microflake | CVT | $1.07 \times 10^{11}$ | 50 | [50] |
| InSe(Gr electrode) | CVT | — | 0.12 | [49] |
| InSe Films | PLD | — | 500 | [13] |
| $Ga_2In_4S_9$ Microflake | CVD | $2.25 \times 10^{11}$ | 40 | [54] |
| $Bi_2O_2Se$ Microflake | CVD | $9.0 \times 10^{13}$ | 0.31 | [55] |
| GaSe Microflake | Bridgman | — | 20 | [56] |
| SnSe Nanowire | CVD | $3.3 \times 10^{12}$ | 0.46 | [47] |
| $In_2Te_3$ Nanowire | CVD | — | 110 | [57] |
| **InSe Nanowire** | **CVD** | $\mathbf{1.57 \times 10^{14}}$ | **0.119** | **This Work** |

## CONCLUSION

In conclusion, we report an edge-homoepitaxial growth method to directly synthesize parallel InSe nanowires on SiO$_2$/Si substrate. This approach enables us to overcome the challenge faced by traditional VLS-governing nanowire growth on amorphous substrate. We propose that the growth mechanism of parallel InSe nanowires is different from conventional VLS approach and may be due to Selenium self-driven vapor-liquid-solid process. Furthermore, we demonstrate that the photodetector based on as-synthesized InSe nanowires simultaneously exhibits a high photoresponsivity (271 A/W), extraordinary specific detectivity ($1.57 \times 10^{14}$ Jones) and fast response speed (rise time = 119 μs, decay time =116 μs). The developed synthesis strategy is compatible with current silicon technology and pave the wave towards for realization of next-generation commercial optoelectronic applications based on nanowires.

## EXPERIMENTAL SECTION

**Material Synthesis:** Indium Selenide (InSe, Sigma-Aldrich, 4N purity) powder was used as the reactant precursors to synthesize InSe microflakes and nanowires on SiO$_2$/Si substrates in a commercial quartz-tube furnace system with a two-temperature zone (Lindberg/Blue M). The high-purity Argon as the carrier gas carried InSe species from the high-temperature zone to the substrate located at the downstream. In a typical procedure of InSe growth, a quartz boat loaded with ~10 mg InSe powders was initially placed at the center of the high-temperature zone, and the other quartz boat loaded with clean SiO$_2$/Si substrates was placed at the downstream of tube about 15-20 cm far away from the high-temperature zone. The tube was pumped to a base pressure of 1 Pa and then flushed by Argon gas flow for 20 mins. During growth of microflakes or nanowires, the rate of gas flow was maintained at 200 and 450 sccm, respectively. The left part of furnace was heated to 800 or 900 °C in 30 mins, and kept for 5 mins, followed by cooling down to 500 °C in 30 mins controlled by a proportion integration differentiation (PID) controller. The right part of furnace was heated to 620 °C in 20 mins, and kept for 30 mins, followed by cooling down to room temperature naturally.

**Material Characterizations:** We performed X-ray photoelectron spectroscopy measurement of as-synthesized sample using PHI VersaProbe 5000 system with Al Kα as X-ray source. The binding energies in this work were calibrated by assigning the corresponding C 1s peak located at 284.5 eV. The compositions and elements distribution of as-synthesized flakes were determined by energy-dispersive X-ray spectroscopy attached to the scanning electron microscope (SEM, Zeiss Gemini 500). We performed Atomic Force Microscopy measurements (Bruker multimode 8) by using ScanAsyst mode. The Raman spectroscopy/mappings were done under a 532.0 nm laser light and silicon-based CCD detector at room temperature using HORIBA JOBIN YVON HR800 Raman system. The spectra have been calibrated by 520.7 cm$^{-1}$ phonon mode of the silicon substrate.

**Sample Transfer and TEM Characterization:** Polymethyl-methacrylate (PMMA) was spin-coated on the as-grown InSe on SiO$_2$/Si substrates, and then PMMA/ InSe stack was transferred onto a TEM grid by etching SiO$_2$ in KOH (2M) solution. Lastly, The PMMA was removed by acetone and isopropanol. High-resolution Transmission Electron Microscopy (TEM, Titan 80-300) and Selected Area Electron Diffraction (SAED) measurements were carried out at the accelerating voltage of 80 kV.

**Fabrication and Measurement of Devices:** The InSe devices were fabricated by a standard electron-beam lithography (FEI F50 with Raith pattern generation system) and electron-beam evaporation (5 nm Ti /45 nm Au). We measured the output and transfer curves of planar devices by using an Agilent B1500A semiconductor analyzer connected to a probe station at room temperature. The optoelectronic measurements of the InSe-based devices were conducted by a Keithley 4200 semiconductor parameter analyzer combined with a lake shore TTPX probe station at room temperature. We performed the time-resolved photocurrent measurements by laser irradiations with square wave modulation and the data are recorded by a digital oscilloscope at a sampling frequency of 1 MHz.

**Mobility Extraction:** The field-effect mobility of device was calculated by fitting the linear region of transfer curve with the following equation:

$$\mu = \frac{L \cdot d}{w \cdot V_{ds} \cdot \varepsilon_0 \cdot \varepsilon_r} \cdot \frac{dI_{ds}}{dV_{bg}}$$

where L and $w$ are the channel length and width, respectively, $\varepsilon_0$ ($\varepsilon_r$) is the vacuum (relative) permittivity, and d is the thickness of $SiO_2$ layer (300 nm).

**Supporting Information**

Photoluminescence/Raman and EDS results of InSe microflake, photoluminescence speactra of bulk InSe and Si substrate, the VSS-growth of InSe nanowires, InSe nanowire device with Schottky contact, the electronic properties of InSe microflake and parallel nanowires, time-dependent OM images and electronic properties of InSe microflakes, Photoresponse performance and Photocurrent trace of InSe microflake photodetector, and scanning photocurrent mapping of the single InSe nanowire photodetector.

**AUTHOR INFORMATION**

**Corresponding Authors**

sjliang@nju.edu.cn;

miao@nju.edu.cn;

**Notes**

The authors declare no competing financial interest

**ACKNOWLEDGEMENTS**

S. Hao and S. N. Yan equally contributed to this work. This work was supported in part by the National Key Basic Research Program of China (2015CB921600), National Natural Science Foundation of China (61625402, 61574076), Shenzhen Basic Research Program (JCYJ20170818110757746), Natural Science Foundation of Jiangsu Province (BK20180330, BK20150055), Fundamental Research Funds for the Central Universities (020414380122, 020414380084), Collaborative Innovation Center of Advanced Microstructures. Dr. S. Hao would like to acknowledge the supports by China Postdoctoral Science Foundation (Grant No 2017M620203) and Postdoctoral Science Foundation of Jiangsu Province.

Received: ((will be filled in by the editorial staff))
Revised: ((will be filled in by the editorial staff))
Published online: ((will be filled in by the editorial staff))
**References**

[1] D. A. Bandurin, A. V. Tyurnina, G. L. Yu, A. Mishchenko, V. Zolyomi, S. V. Morozov, R. K. Kumar, R. V. Gorbachev, Z. R. Kudrynskyi, S. Pezzini, Z. D. Kovalyuk, U. Zeitler, K. S. Novoselov, A. Patane, L. Eaves, I. V. Grigorieva, V. I. Fal'ko, A. K. Geim, Y. Cao, *Nat. Nanotechnol*. **2017**, *12*, 223.

[2] J. Zeng, S.-J. Liang, A. Gao, Y. Wang, C. Pan, C. Wu, E. Liu, L. Zhang, T. Cao, X. Liu, Y. Fu, Y. Wang, K. Watanabe, T. Taniguchi, H. Lu, F. Miao, *Phys. Rev. B* **2018**, *98,* 125414.

[3] M. Li, C. Y. Lin, S. H. Yang, Y. M. Chang, J. K. Chang, F. S. Yang, C. Zhong, W. B. Jian, C. H. Lien, C. H. Ho, H. J. Liu, R. Huang, W. Li, Y. F. Lin, J. Chu, *Adv. Mater*. **2018**, *30*, e1803690.

[4] S. R. Tamalampudi, Y. Y. Lu, U. R. Kumar, R. Sankar, C. D. Liao, B. K. Moorthy, C. H. Cheng, F. C. Chou, Y. T. Chen, *Nano Lett.* **2014**, *14*, 2800.

[5] S. Lei, F. Wen, L. Ge, S. Najmaei, A. George, Y. Gong, W. Gao, Z. Jin, B. Li, J. Lou, J. Kono, R. Vajtai, P. Ajayan, N. J. Halas, *Nano Lett.* **2015**, *15*, 3048.

[6] Z. Li, H. Qiao, Z. Guo, X. Ren, Z. Huang, X. Qi, S. C. Dhanabalan, J. S. Ponraj, D. Zhang, J. Li, J. Zhao, J. Zhong, H. Zhang, *Adv. Funct. Mater.* **2018**, *28*, 1705237.

[7] J.-J. Wang, F.-F. Cao, L. Jiang, Y.-G. Guo, W.-P. Hu, L.-J. Wan, *J. Am. Chem. Soc.* **2009**, *131*, 15602.

[8] S. Lei, L. Ge, S. Najmaei, A. George, R. Kappera, J. Lou, M. Chhowalla, H. Yamaguchi, G. Gupta, R. Vajtai, A. D. Mohite, P. M. Ajayan, *ACS Nano* **2014**, *8*, 1263.

[9] W. Zhou, G. Yu, A. N. Rudenko, S. Yuan, *Phys. Rev. Mater.* **2018**, *2,* 114001.

[10] M. Wu, J. J. Shi, M. Zhang, Y. M. Ding, H. Wang, Y. L. Cen, W. H. Guo, S. H. Pan, Y. H. Zhu, *Nanotechnology* **2018**, *29*, 205708.

[11] J. Zhou, J. Shi, Q. Zeng, Y. Chen, L. Niu, F. Liu, T. Yu, K. Suenaga, X. Liu, J. Lin, Z. Liu, *2D Mater*. **2018**, *5*, 025019.

[12] H. C. Chang, C. L. Tu, K. I. Lin, J. Pu, T. Takenobu, C. N. Hsiao, C. H. Chen, *Small* **2018**, *14*, e1802351.

[13] Z. Yang, W. Jie, C. H. Mak, S. Lin, H. Lin, X. Yang, F. Yan, S. P. Lau, J. Hao, *ACS*
16